\documentclass[mathleft]{an}
\usepackage{graphicx}
\usepackage{times}
\usepackage{subfigure}
\def\Msun{\hbox{$\thinspace M_{\odot}$}}

\def\xmm{XMM-{\it Newton}}

\def\N1316{NGC\,1316}
\def\N1404{NGC\,1404}

\def\4U{4U~1735$-$444}

\def\xmm{XMM-{\it Newton}}

\def\arcsec{\ifmmode '' \else $''$\fi}

\def\arcsecpoint{\ifmmode ''\!. \else $''\!.$\fi}

\def\kms{\ifmmode {\rm km\ s}^{-1} \else km s$^{-1}$\fi}
\def\Msun{\ifmmode {\rm M}_{\odot} \else M$_{\odot}$\fi}
\def\Lsun{\ifmmode {\rm L}_{\odot} \else L$_{\odot}$\fi}
\def\Zsun{\ifmmode {\rm Z}_{\odot} \else Z$_{\odot}$\fi}

\def\ergscm2{ergs\,s$^{-1}$\,cm$^{-2}$}
\def\icm3{{\rm cm}^{-3}}
\def\icm2{{\rm cm}^{-2}}
\def\qo{\ifmmode q_{\rm o} \else $q_{\rm o}$\fi}
\def\Ho{\ifmmode H_{\rm o} \else $H_{\rm o}$\fi}
\def\ho{\ifmmode h_{\rm o} \else $h_{\rm o}$\fi}

\def\vFWHM{\ifmmode v_{\mbox{\tiny FWHM}} \else
            $v_{\mbox{\tiny FWHM}}$\fi}
\def\CCF{\ifmmode F_{\it CCF} \else $F_{\it CCF}$\fi}
\def\ACF{\ifmmode F_{\it ACF} \else $F_{\it ACF}$\fi}
\def\Halpha{\ifmmode {\rm H}\alpha \else H$\alpha$\fi}
\def\Hbeta{\ifmmode {\rm H}\beta \else H$\beta$\fi}
\def\Hgamma{\ifmmode {\rm H}\gamma \else H$\gamma$\fi}
\def\Hdelta{\ifmmode {\rm H}\delta \else H$\delta$\fi}
\def\Lya{\ifmmode {\rm Ly}\alpha \else Ly$\alpha$\fi}
\def\Lyb{\ifmmode {\rm Ly}\beta \else Ly$\beta$\fi}
\def\Lyg{\ifmmode {\rm Ly}\beta \else Ly$\gamma$\fi}

\def\ciii{\ifmmode {\rm C}\,{\sc iii} \else C\,{\sc iii}\fi}
\def\civ{\ifmmode {\rm C}\,{\sc iv} \else C\,{\sc iv}\fi}
\def\cv{\ifmmode {\rm C}\,{\sc v} \else C\,{\sc v}\fi}
\def\cvi{\ifmmode {\rm C}\,{\sc vi} \else C\,{\sc vi}\fi}

\def\o5007{[O\,{\sc iii}]\,$\lambda5007$}

\def\ovii{O\,{\sc vii}}
\def\oviii{O\,{\sc viii}}

\def\nex{Ne\,{\sc x}}

\def\fexvii{Fe\,{\sc xvii}}

\def\xmm{XMM-{\it Newton}}

\def\gsim{\mathrel{\hbox{\rlap{\hbox{\lower4pt\hbox{$\sim$}}}\hbox{$>$}}}}
\def\lsim{\mathrel{\hbox{\rlap{\hbox{\lower4pt\hbox{$\sim$}}}\hbox{$<$}}}}

\begin{document}

\Pagespan{000}{}
\Yearpublication{0000}%
\Yearsubmission{0000}%
\Month{00}%
\Volume{000}%
\Issue{00}%
\DOI{This.is/not.aDOI}%


\title{Ultrafast outflows in ultraluminous X-ray sources}

\author{Ciro Pinto\inst{1}\fnmsep\thanks{Corresponding author:
  \email{cpinto@ast.cam.ac.uk}\newline}
\and  ~Andy Fabian~\inst{1}
\and  ~Matthew Middleton\inst{1,2}
\and  Dom Walton\inst{1,3}
}
\titlerunning{Instructions for authors}
\authorrunning{Pinto et al }
\institute{
Institute of Astronomy, Madingley Rd, Cambridge, CB3 0HA, UK
\and 
Physics \& Astronomy, University of Southampton, Southampton, Hampshire SO17 1BJ, UK
\and 
Jet Propulsion Laboratory, California Institute of Technology, Pasadena, CA 91109, USA
}
\received{\today}


\abstract{%
Ultraluminous X-ray sources (ULXs) are bright extragalactic sources with X-ray luminosities
above $10^{39}$\,erg/s powered by accretion onto compact objects.
According to the first studies performed with XMM-\textit{Newton} ULXs seemed to be excellent candidates 
to host intermediate-mass black holes $(10^{2-4} M_{\odot})$. 
However, in the last years the interpretation of super-Eddington accretion onto stellar-mass black holes 
or neutron stars for most ULXs has gained a strong consensus.
One critical missing piece to confirm the super-Eddington scenario was the direct detection of the
massive, radiatively-driven winds expected as atomic emission/absorption lines in ULX spectra. 
The first evidence for winds was found as residuals in the soft X-ray spectra of ULXs.
Most recently we have been able to resolve these residuals into rest-frame emission and 
blueshifted ($\sim0.2c$) absorption lines arising from highly ionized gas 
in the deep high-resolution XMM-\textit{Newton} spectra of two ultraluminous X-ray sources. 
The compact object is therefore surrounded by powerful ultrafast winds as predicted
by models of hyper-Eddington accretion. Here we discuss the relevance of these discoveries
and the importance of further, deep, XMM-\textit{Newton} observations of powerful winds 
in many other ultraluminous X-ray sources to estimate the energetics of the wind, 
the geometry of the system, and the masses of the central accretors.}

\maketitle

\section{Introduction}

Ultraluminous X-ray sources (ULXs) are extragalactic, off-nucleus, point sources in galaxies
and have X-ray luminosities that exceed the Eddington limit for a 10\Msun black hole (BH), or $10^{39}$\,erg/s. 
Early detections of ULXs were possible with the \textit{Einstein} observatory
(Long \& van Speybroeck 1983). In the last two decades, X-ray surveys have observed hundreds of ULXs
with different facilities (ROSAT, Liu \& Bregman 2005; \textit{Chandra}, Swartz et al. 2004; XMM-\textit{Newton},
Walton et al. 2011). Two main scenarios have been proposed to describe 
the nature of these exotic objects, both implying accretion onto compact objects.

One scenario proposes ULXs to be powered by intermediate-mass $(10^{2-4} M_{\odot})$ black holes (IMBHs)
accreting in a sub-Eddington regime similar to the well-studied Galactic BHs (e.g., Kaaret et al. 2001; 
Miller et al. 2003). Probing the census of IMBHs is a crucial ingredient to understand the growth 
of supermassive black holes (SMBHs) powering active galaxies (Kormendy \& Ho 2013).
These objects are also the most likely explanation for the far bright end of the ULXs, 
called hyperluminous X-ray sources, whose prototype is ESO 243-49 HLX-1 (Farrell et al. 2009).
Its outburst and spectral states are similar to the sub-Eddington Galactic BHs.

The other scenario describes ULXs as stellar-mass black holes or neutron stars accreting 
several orders of magnitude above the classic Eddington limit (super-Eddington or super-critical
accretion). Super-Eddington accretion can be explained with several models:
photon-bubble instability in thin disks (Begelman 2002), inefficient regime of accretion in slim disks
(Kawaguchi 2003), and radiation-supported thick disks (Poutanen et al. 2007).
Geometrical beaming (King et al. 2001) may provide a further increase of luminosity 
in the line of sight, although observations rule out extreme beaming in ULXs 
{(e.g., Berghea et al. 2010; Moon et al. 2011)}.
Super-Eddington accretion is another crucial ingredient to understand SMBHs growth 
because the discovery of fully grown SMBH in early stages of the Universe requires 
long periods of high accretion (Fan et al. 2003, Volonteri et al. 2013).

ULXs enable both the study of super-Eddington accretion and the search for 
IMBHs and, therefore, have an important impact on cosmology.

The first masses inferred dynamically from optical data revealed two ULXs 
to be powered by compact objects with masses in the stellar-mass black hole range 
(Liu et al. 2013; Motch et al. 2014). The detection of bright radio emission confirmed
another stellar-mass compact object in a transient source in M\,31 (Middleton et al. 2013).
M\,82 X-2 shows pulsations, which indicates a neutron star as the accretor (Bachetti et al. 2014).
A clear downturn has been detected in most high-quality XMM-\textit{Newton} and NuSTAR ULX spectra 
at energies of a few keV (e.g. Roberts et al. 2005; Stobbart, Roberts \& Wilms 2006; 
Bachetti et al. 2013; Walton et al. 2014). 
Most ULX X-ray spectra can therefore be described by
a combination of a hard component showing this downturn
and a soft excess, which is not consistent with the classic sub-Eddington BH accretion states.

In a most recent work, Pinto et al. (2016), 
we have discovered the presence of rest-frame emission 
and blueshifted (∼0.2$c$) absorption lines arising from highly ionized gas 
in the deep high-resolution XMM-\textit{Newton} spectra of two ultraluminous X-ray sources:
NGC 1313 X-1 and NGC 5408 X-1. These features resolved the soft X-ray residuals 
that were earlier detected in the spectra of these and other ULXs taken with the 
European Photon Imaging Camera (EPIC, CCD) at moderate resolution (Middleton et al. 2014, 2015).
This discovery was further strengthened by a follow up paper in which we detected
its highly energetic Fe K counterpart in the long-exposure XMM-\textit{Newton} CCD
and NuSTAR CdZnTe spectra (Walton et al. 2016).
The discovery of the long sought powerful winds in ULXs provides further support
to the scenario of super-Eddington accretion and
in this paper we discuss the relevance of these discoveries
and the importance of further, deep, XMM-\textit{Newton} observations of powerful winds 
in many other ultraluminous X-ray sources in order to estimate the energetics of the wind, 
the geometry of the system, and the black hole masses.

\section{Soft X-ray spectral features of ULXs}

It has been known for a decade that soft ULXs show residuals to their continuum
spectral fits (e.g. Stobbart et al. 2006). It was not clear whether they
were produced by the hot phase of the interstellar medium of the host galaxies, 
by gas shocks in the optical nebulae around the ULXs 
or by the gas accreted onto the compact objects.
Middleton et al. (2014) investigated the residuals in the archetypal ULXs
NGC 5408 X-1 and NGC 6946 X-1, and found that
they can be explained as broad absorption lines from a partially
ionised and blueshifted $(v\sim0.1c)$ medium in agreement
with absorption from the optically thin phase of a turbulent
outflowing wind, located outside the last scattering
surface of the optically-thick region.
Sutton, Roberts, \& Middleton (2015) used \textit{Chandra} high spatial-resolution
spectra of NGC 5408 X-1 to show that the residuals were mostly associated with the 
ULX itself rather than to star formation or hot gas in the host galaxy.

\subsection{The shape of the spectral residuals}

\begin{figure}
\centering
\includegraphics[width=0.475\textwidth,angle=0]{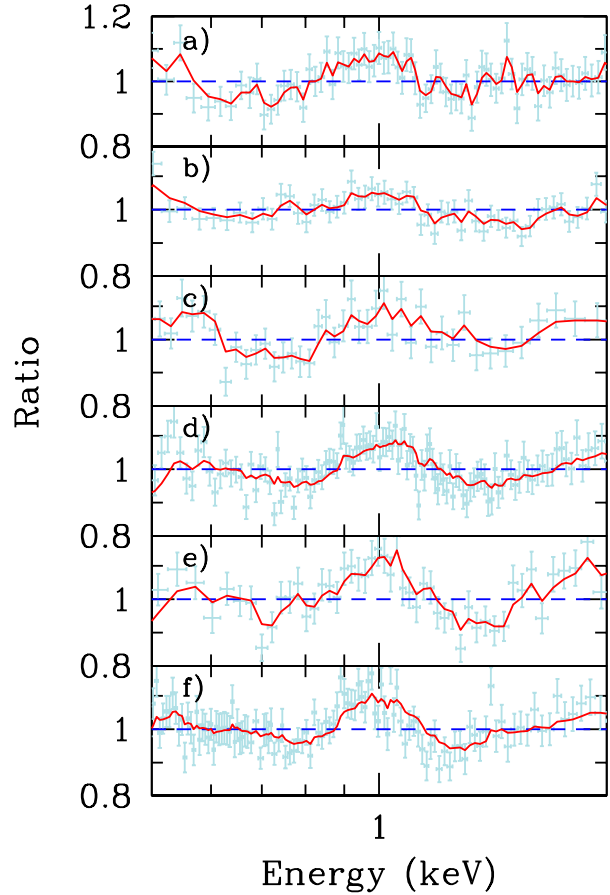}
\caption{Soft X-ray residuals to continuum models for 6 ULXs.
The objects are: (a) NGC 1313 X-1, (b) Ho IX X-1, (c) Ho II X-1,
(d) NGC 55 ULX-1, (e) NGC 6946 X-1, and (f) NGC 5408 X-1.
The spectral data are shown in light blue and plotted as a ratio to
the best fitting model, and an exponentially smoothed function is
plotted in red to highlight the commonality in the shape of the
residuals. Figure from Middleton et al. (2015).}
\label{fig:Middleton15fig1}
\end{figure}

\begin{figure}
\centering
\includegraphics[width=0.475\textwidth,angle=0]{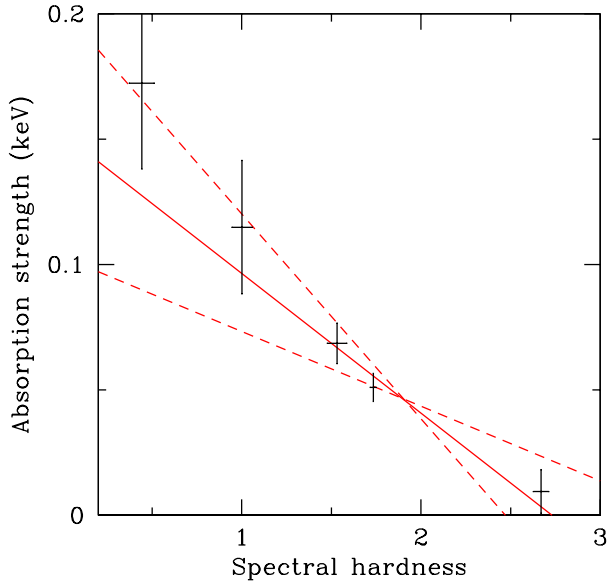}
\caption{Absorption strength from the spectral fits plotted against
spectral hardness (with 1$\sigma$ error bars). The solid red line is the best-fitting
relation (with a slope of $−0.05\pm0.01$) and the dashed lines are the
extremal range of relations (from the 3$\sigma$ errors on the slope
and intercept). Figure from Middleton et al. (2015).}
\label{fig:Middleton15fig2}
\end{figure}

In a crucial step towards the understanding of the spectral residuals
we show that their shape is similar among six different ULXs with high
quality soft X-ray spectra (Middleton et al. 2015, see also Fig.\,\ref{fig:Middleton15fig1}),
which suggests that they have the same origin.
We have further probed the relation between the residuals 
and the spectral states of NGC 1313 X-1, which has the largest dynamical range in
its hardness of any of the ULXs with detectable residuals.
A clear anti-correlation between spectral hardness 
(ratio between the flux in the 1--10\,keV and the 0.3--1\,keV bands) 
and the strength (or equivalent width) of the features
was found (see Fig.\,\ref{fig:Middleton15fig2}), which rules out 
an origin due to reflection on top of the accretion disk (expected 
to increase with the hardness) or to any galactic / star formation event.
This is instead consistent with wind models, where for moderate-to-high
inclination angles an increase in accretion rate leads
to a stronger wind, which both softens the spectrum and
deposits more material into the optically thin medium surrounding
the ULX, that deepens the absorption features (e.g., Poutanen et al. 2007).

\subsection{High-resolution X-ray spectroscopy with gratings}

The XMM-\textit{Newton} and \textit{Chandra} X-ray satellites are provided with spectrometers
that have a high $(E / \Delta E \sim 100 - 1000)$ energy spectral resolution and are, 
therefore, able to resolved spectral features narrower than $1000$\,km\,s$^{-1}$: 
the gratings (Brinkman et al. 2000, den Herder et al. 2001 and Canizares et al. 2005).
Despite their high resolution power, the gratings are often ignored because
they have much lower effective area than the CCD spectrometers, 
provide only 1D spectroscopy (along their cross dispersion direction)
and require a careful data reduction, particularly for extended and/or multiple sources 
in the field of view. However, at least for XMM-\textit{Newton},
the reflection grating spectrometers (RGS) work in parallel with the EPIC cameras,
and the archive is therefore rich with unexplored high-spectral resolution RGS data.
There is a wealth of RGS public observations of ULXs: NGC 1313 X-1 and NGC 5408 X-1, 
in particular, were observed for several XMM-\textit{Newton} orbits.

\subsection{Zooming onto NGC\,1313\,X-1 and NGC\,5408\,X-1}

NGC 1313 X-1 and NGC 5408 X-1 are two archetypal ULXs showing all the classical properties
of the bulk of the population and have X-ray (0.2--10\,keV) luminosities up to $\sim10^{40}$\,erg/s.
They have been extensively observed in spectral states where a large fraction of their flux emerges 
below 2 keV (Stobbart et al. 2006; Gladstone et al. 2009)
showing strong spectral deviations (0.6--1.2 keV) from the underlying continuum 
in CCD spectra (see Fig.\,\ref{fig:Middleton15fig1}). 
These two ULXs are within a 5\,Mpc distance, are rather isolated, and 
have the longest set of full XMM-\textit{Newton} orbits. This allows for a comparatively clean, 
high energy-resolution study of the features seen in CCD spectra.

\begin{figure*}
\centering
\mbox{\subfigure{\includegraphics[width=0.45\textwidth,angle=0]{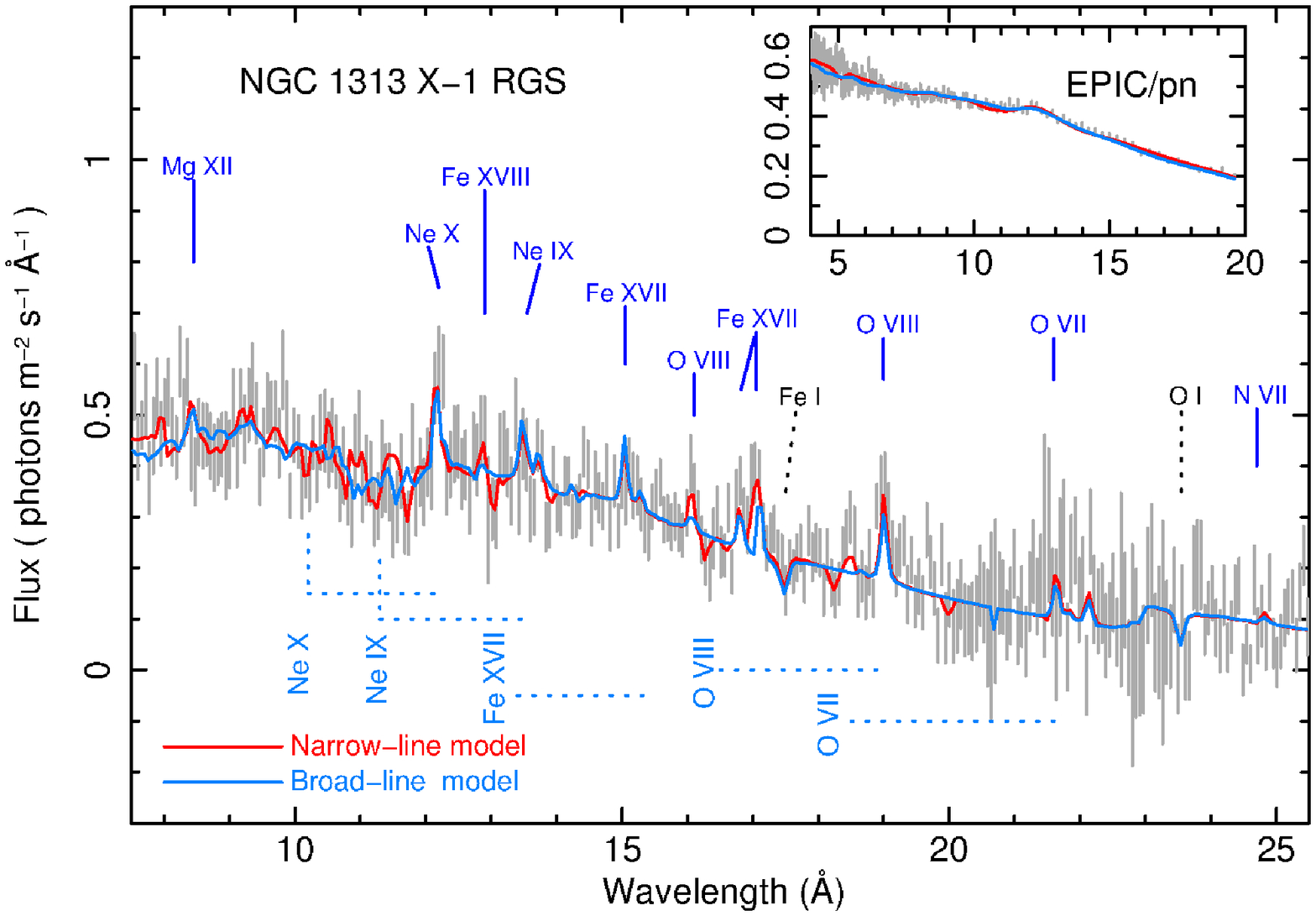}}
\hspace{15pt}
\subfigure{\includegraphics[width=0.45\textwidth,angle=0]{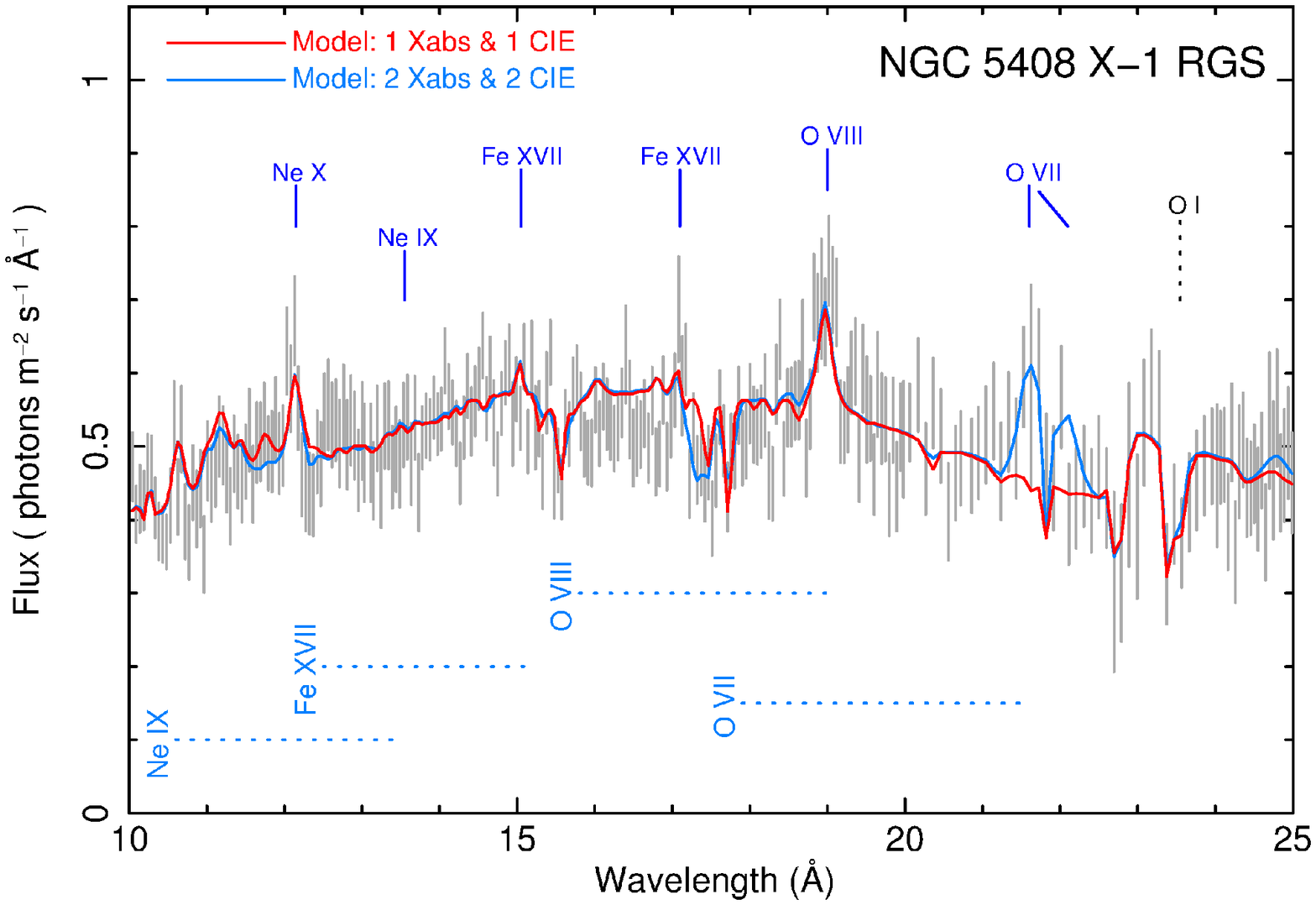}}
}\\
\mbox{\subfigure{\includegraphics[width=0.45\textwidth,angle=0]{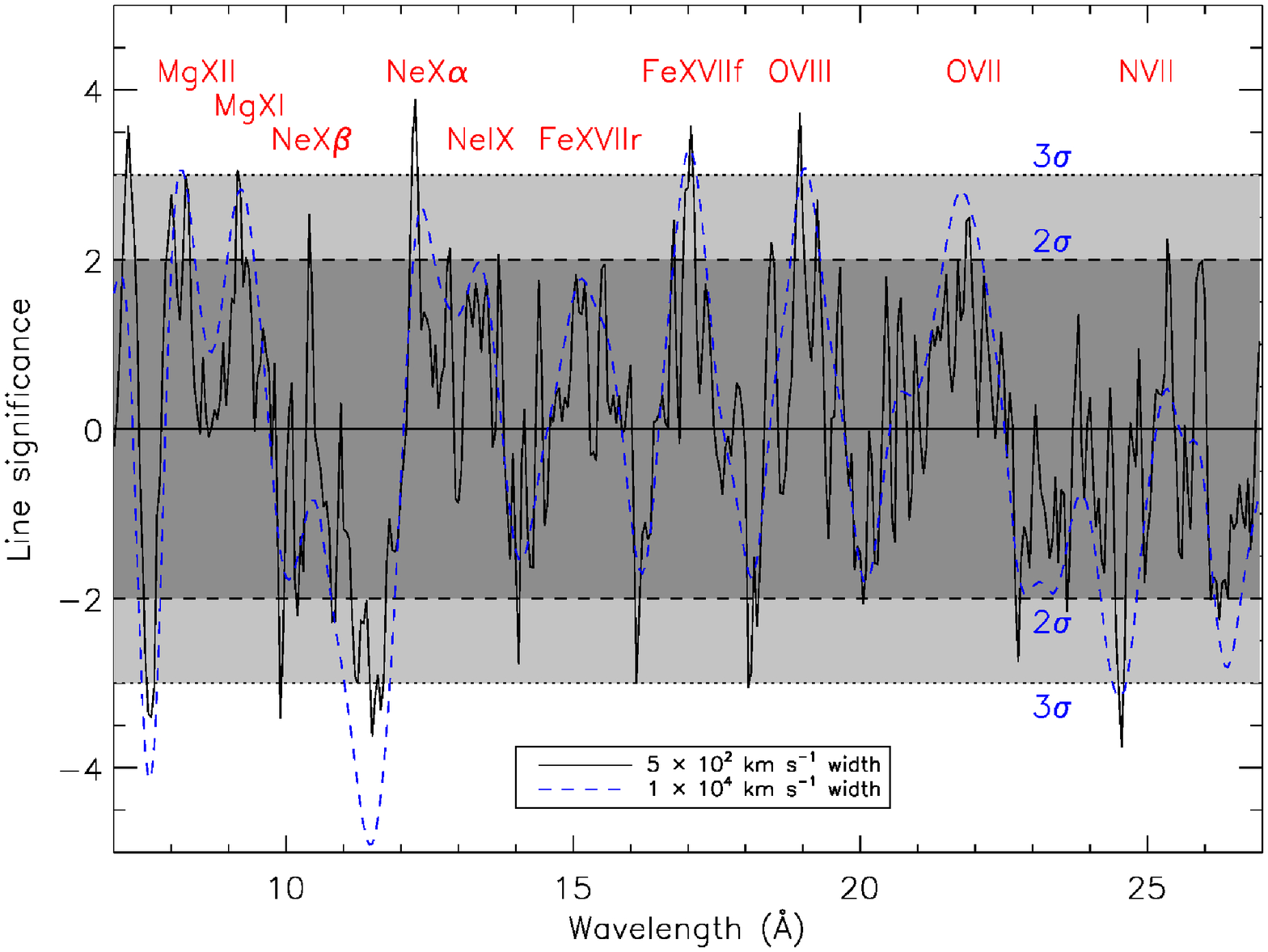}}
\hspace{15pt}
\subfigure{\includegraphics[width=0.45\textwidth,angle=0]{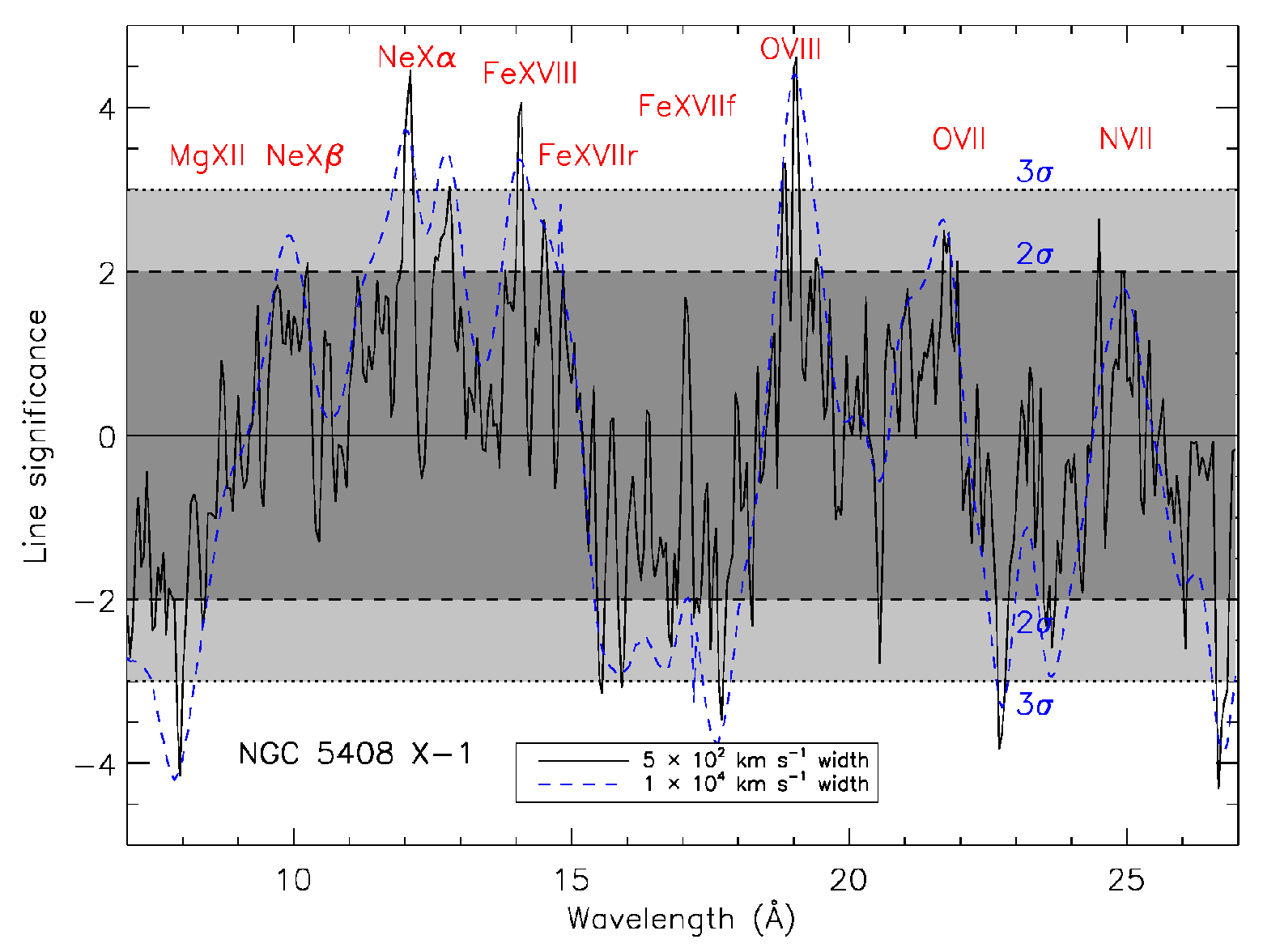}}
}\\
\caption{XMM-Newton/RGS high-resolution grating spectra of NGC 1313 X-1 and NGC 5408 X-1 (top)
with overlaid a model of thermal collisionally-ionized emission and a relativistically 
$(v\sim0.2c$ blueshifted photoionized absorption. The dotted lines indicate the blueshift 
from the rest-frame transitions (identical for all absorption lines).
Bottom plots show the corresponding line significance obtained by Gaussian fits
with increments of 0.05 {\AA} and negative values indicating absorption lines. 
All plots are from Pinto et al. (2016).} 
\label{fig:Pinto16fig1}
\end{figure*}

In Pinto et al. (2016) we collect the three and six full XMM-\textit{Newton} orbits
of NGC 1313 X-1 and NGC 5408 X-1, obtaining 345.6\,ks and 644.9\,ks solar-flared clean data.
We stack the spectra of each source with an advanced technique that accounts for the different 
background subtraction of the two RGS spectrometers. 
This provided the two best high-resolution X-ray spectra 
ever taken for ULXs (see Fig.\ref{fig:Pinto16fig1}).

The RGS spectra of NGC 1313 X-1 and NGC 5408 X-1 show a wealth of emission and absorption features
including the absorption features produced by the foreground Galactic interstellar medium 
(see, e.g., Pinto et al. 2013). At first, we could identify strong, rest-frame emission lines 
from a mixture of elements at varying degrees of ionization, including \nex\,($\lambda=12.1$\,{\AA}),
\oviii\,($19.0$\,{\AA}), \ovii\,($21.6$\,{\AA}), and \fexvii\,($15.0,17.1$\,{\AA}).
There are several narrow absorption-like features including a broad depression at the blue side
($10-12$\,{\AA}) of the \nex\ emission line in NGC 1313 X-1. 
These spectral features actually resolve the residuals shown by Middleton et al. (2014, 2015).
Before trying to model these emission and absorption
features we constrain the spectral continuum by simultaneously fitting the EPIC and RGS spectra
of these sources with a standard blackbody plus powerlaw emission model,
multiplied by neutral absorption both from our Galaxy and intrinsic to NGC 1313.

We analyze the spectral features using the high-resolution RGS spectra
and adopting the combined EPIC--RGS spectral continuum.
We first measure their significance by fitting a Gaussian over
the 7--27\,{\AA} wavelength range with increments of 0.05\,{\AA}. 
We have also tested the effects of adopting different line widths.
The strongest emission lines, found at their rest frame wavelengths,
are detected at a $3\sigma$ level each. The strongest absorption features, 
e.g. between $10-12$\,{\AA} in NGC 1313 X-1, are also detected at $3\sigma$,
if interpreted as blueshifted {Ne\,{\sc ix-x}} or {Fe\,{\sc xviii}} absorption
(and accounting for the look-elsewhere effect).

\subsection{Ultrafast outflows in the ULX grating spectra}

We modeled the emission lines with a rest-frame, collisional ionization equilibrium (CIE)
plasma, which includes an underlying weak thermal continuum
at an average temperature of 0.8 keV ($\sim10^7$\, K; see Fig.\ref{fig:Pinto16fig1}).
The absorption lines can be well modelled with a mildly ionized 
($\log\,\xi=2.25\pm0.05$ or $\xi\sim$ 180\,erg\,cm\,s$^{-1}$)
absorbing gas in photoionization equilibrium applied to the continuum.
This absorber has a tremendous outflow velocity of $\sim0.2$ times the speed of light.
The inclusion of a rest-frame absorber with the same ionization parameter 
provides a slight improvement to the fit. These model adopts the same line width 
for absorption and emission lines ($<$\,500\,km\,s$^{-1}$).
The large depression between $10-12$\,{\AA} in NGC 1313 X-1
can also be modeled with an absorption with higher ($\log\,\xi=4.5$) ionization parameter 
and relativistic ($\sim0.1c$) line broadening 
(see blue line in Fig.\ref{fig:Pinto16fig1}, top left panel).
NGC 5408 X-1 also shows sharp emission features similar to those
detected in NGC 1313 X-1, but the outflow seems to be more structured
with the emission lines being stronger than the absorption features.
A better fit is obtained with a multiphase emission (CIE) and absorption (Xabs) model
(see blue line in Fig.\ref{fig:Pinto16fig1}, top right panel).
This could suggest a different geometry or view angle for the two ULXs.
More detail about the spectral modeling of NGC 1313 X-1 and NGC 5408 X-1 can be
found in Pinto et al. (2016). 

This is the fastest wind ever detected in a X-ray binary and it could carry
enough power to affect its neighbourhood. The outflow rate can be written as
{\.{M}}\,$ = 4 \pi R^2 \rho v \Omega$, which gives a wind power 
$P_w = 1/2$\,{\.{M}}\,$v^2 = 2 \pi R^2 m_{\rm p} n_{\rm H} v^3 \Omega$, 
where $m_{\rm p}$ is the proton mass and $\Omega$ the solid angle. 
Since, $\xi = L/n_{\rm H}R^2$, using the results for the relativistically-
broadened absorber, we measure an upper limit for the 
ratio between the wind power and the luminosity
of NGC 1313 X-1 to be $P_w/L = 100\,\Omega\,C_V$. 
This would imply a highly super-Eddington accretion rate, 
unless both the covering fraction $C_V$ and the duty cycle are an order of magnitude
lower than unity, which is not the case for most ULXs.
NGC 1313 X-1, for instance, shows significant spectral features 
for a large range of spectral hardness (Middleton et al. 2015),
which suggests a large covering fraction.

\subsection{Fe\,K powerful ultrafast outflow in NGC 1313 X-1}

Motivated by the discovery of a relativistic outflow in these soft X-ray spectra of ULXs,
we used the EPIC-pn\,/\,MOS\,1-2 and FPMA\,/\,B longest exposures from XMM-\textit{Newton}
and NuSTAR, respectively. These provided total good exposures of 258 ks for EPIC-pn,
331 ks for each EPIC-MOS and 360 ks per FPM (see Walton et al. 2016). 
We simultaneously fitted these spectra with a cutoff powerlaw (e.g., Stobbart et al. 2006)
including neutral absorption both from our Galaxy and intrinsic to NGC 1313.

We searched for Fe\,K absorption or emission following Walton et al. (2012): 
we included a narrow (intrinsic width of $\sigma = 10$ eV) Gaussian, 
and varied its energy across the energy range of interest in steps of 40 eV
(oversampling the XMM-Newton energy resolution by a factor $\sim 4-5$). 
The Gaussian normalisation can be either positive
or negative. For each line energy, we record the $\Delta \chi^2$ 
improvement in fit resulting from the inclusion of the Gaussian
line, as well as the best fit equivalent width $(EW)$ and its
90 (1.64$\sigma$) and 99\% (2.58$\sigma$) confidence limits. These are calculated with the
EQWIDTH command in XSPEC, using 10,000 parameter simulations 
based on the best fit model parameters and their uncertainties. 
To be conservative, we vary the Gaussian line energy
between 6.6 and 9.6 keV, corresponding to a wide range of
outflow velocities extending up to $>0.25c$ for {Fe\,{\sc xxvi}}.

\begin{figure}
\centering
\includegraphics[width=0.45\textwidth,angle=0]{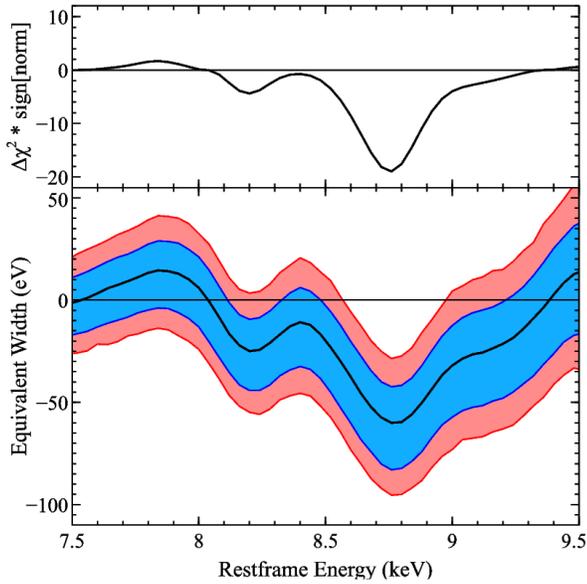}
\caption{Detection of Fe K blueshifted absorption (6.70\,keV lab energy)
in the combined XMM-\textit{Newton} EPIC CCD and NuSTAR FPMA/B CdZnTe spectra. 
Figure from Walton et al. (2016).}
\label{fig:Walton15fig1}
\end{figure}

The results are shown in Fig.\,\ref{fig:Walton15fig1} (see also Walton et al. 2016).
We detected an absorption line at $8.77\pm^{0.05}_{0.06}$\,keV with
an equivalent width of $-61\pm24$eV, which is comparable 
to the strongest iron absorption seen from a black hold binary to date 
(King et al. 2012). 
We successfully modeled the feature with a photoionized absorber. 
The best-fit provides ionization parameter $\log \xi = 3.3\pm^{0.3}_{0.5}$, 
which is higher than that of the gas we detected in the soft X-ray band,
but its outflow velocity of $v_{\rm out} = 0.236 \pm 0.005c$ 
is in very good agreement with the low-ionization soft X-ray absorber.
This fit adopts {Fe\,{\sc xxv}} as dominant ion.
Another fit solution with {Fe\,{\sc xxvi}} leading the iron ionic distribution
and $v_{\rm out}\sim0.2c$ is not excluded, but still confirming
the presence of a relativistic outflow.
We measure a ratio between the wind power and the bolometric luminosity of 
$P_w/L = 1500\,(60)\,\Omega\,C_V$ for the ionization parameter $\log \xi = 3.3$ (4.5),
suggesting that the wind may dominate the energy output from NGC 1313 X-1.
Regardless of the precise value of $\Omega$ and $C_V$, these measurements
are absolutely extreme if compared to typical sub-Eddington accreting 
black holes and agree with the predictions for a super-Eddington
accretion scenario (e.g., Poutanen et al. 2007; King 2010).

\section{Discussion}

\subsection{{\xmm} -- NuSTAR joint efforts}

NuSTAR observations have clearly shown that the spectral states 
of typical ultraluminous X-ray sources present a cutoff above 10 keV, 
which rules out the interpretation in terms of black holes in a standard
low/hard state or reflection-dominated disk regime.
Several ULXs, e.g. NGC 1313 X-2, also show spectral transition from
states of high luminosity and variability to states of low luminosity
without strong variability as expected in transitions from a super-Eddington 
to a sub-Eddington regime (e.g., Bachetti et al. 2013).
The detection of powerful winds in ultraluminous X-ray sources
was a missing ingredient for the super-Eddington interpretation
of their X-ray luminosities, but our important discoveries of relativistic winds
have filled this gap with the theoretical predictions.

\begin{figure}
\centering
\includegraphics[width=0.485\textwidth,angle=0]{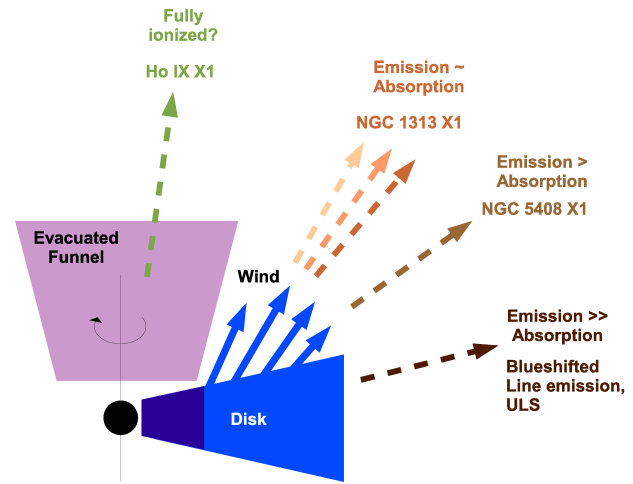}
\caption{Cartoon of a physical model of high mass
accretion rate sources. The light blue region shows the
soft X-ray emission of the accretion disk, altered by a
photosphere of a radiatively-driven optically-thick
wind. The dark blue region, closer to the compact
object is dominated by highly variable, optically-thin,
turbulent Comptonization emitting high-energy ($>1$ keV) X-rays.}
\label{fig:ULX_geometry}
\end{figure}

\subsection{Further insights in super-Eddington accretion}

At high accretion rates the disk thickens due to radiation pressure and 
cools via advection and/or winds; the bolometric luminosity exceeds the 
Eddington luminosity by a factor proportional to $\log$ \.m (Poutanen et al. 2007). 
The accretion can increase, orders of magnitude larger than the Eddington limit, 
where the huge radiation pushes out part of the disk atmosphere at extreme velocities, 
giving the system a geometry similar to that of a funnel (Fig.\,\ref{fig:ULX_geometry}). 
The funnel can significantly collimate the X-ray emission via scattering, 
in which case the disk brightens further (geometrical beaming). 
The agreement between theory and spectral variability suggests that 
ULXs are super-Eddington accreting, stellar mass BHs similar to SS 433,
but viewed close to the symmetric axis (e.g., King et al. 2002; Fabrika 2004). 

\subsection{Wind dependence on the line of sight}

In Middleton et al. (2014, 2015) we used the moderate-resolution, 
but high-sensitivity CCDs on board {\xmm} to show that the absorption features 
anti-correlate with the spectral hardness. 
We believe that this trend depends on the viewing angle (see Fig.\,\ref{fig:ULX_geometry}).
Close to the symmetric axis the hard X-rays dominate the spectrum smearing out the wind features (Ho IX X-1). 
At intermediate angles we look directly through the wind and detect strong absorption features, which can vary
due to precession (NGC 1313 X-1). At large angles we only see a small portion of the disk; 
the emission lines, coming from a larger region, are stronger than the absorption lines (NGC 5408 X-1). 
Edge on, the inner regions are completely obscured by a large amount of cold, 
optically-thick gas; the source peaks in UV looking like an ultraluminous supersoft X-ray source (Liu et al. 2015). 
This scenario gives an elegant explanation to the vast phenomenology of ULXs, 
but needs further evidence with a large sample of ULXs.

\subsection{ULXs as cosmological probes}

The detection of fully grown SMBHs at high redshifts, and therefore in early stages of the Universe,
challenge the theories of the classical sub-Eddington accretion, requiring 
long periods of higher accretion (see, e.g., Fan et al. 2003, Volonteri et al. 2013).
Thorough studies of accretion in high-redshift AGN is beyond the capabilities 
of the current mission. ULXs are bright, nearby super-Eddington accretors
and therefore provide the best means to study the phenomenology of extreme accretors.

\subsection{XMM legacy of super-Eddington accretors}

Undoubtedly, a statistical sample of typical ULXs is indispensable to confirm
the super-Eddington scenario in these sources and to study it in detail.
This requires to collect deep gratings and CCD spectra of other ULXs
in order to resolve the soft X-ray ubiquitously detected in CCD spectra 
(see Fig.\,\ref{fig:Middleton15fig1}) as well as to probe the highly-ionized 
Fe\,K part of the wind. 
Here we have used NGC\,1313\,X-1 and NGC\,5408\,X-1, two archetypal bright 
($L_X\sim10^{40}$\,erg/s) ULX, due to their large $>350$\,ks {\xmm} exposures
and we have obtained excellent results, which are encouraging for the future.
There are at least a dozen of ULXs with a comparable luminosity and flux
that can be thoroughly studied, including those in Fig.\ref{fig:Middleton15fig1}
(Swartz et al. 2004, Liu \& Bregman 2005; Walton et al. 2011).
We estimated that $\sim2$ full {\xmm} orbits per ULX 
will allow us to resolve the residuals in their EPIC spectra
similar to the two ULXs studied here (accounting for solar flares).
The gratings on board \textit{Chandra} will also provide an excellent
workbench to detect and analyze winds in ULX; in particular
they may show some exciting spectral features in the $1.8-6$\,keV energy band
where the {\xmm} instruments lack the require spectral resolution.
Due to their lower effective area, the \textit{Chandra} gratings 
will require exposures of at least 500\,ks per ULX.




\acknowledgements
CP and ACF acknowledge support from ERC Advanced Grant 340442.  
MJM acknowledges support from an STFC Ernest Rutherford fellowship.
This paper is based on observations obtained with \xmm, 
an ESA science mission with instruments and contributions 
directly funded by ESA Member States and the USA (NASA),
and NuSTAR, a project led by Caltech, funded by NASA and
managed by NASA/JPL.   



\end{document}